\documentclass[pra,showpacs,twocolumn,superscriptaddress]{revtex4}

\usepackage{graphicx}
\usepackage{bbm}
\usepackage{amsmath}
\usepackage{amsfonts}
\usepackage{amssymb}

\newcommand{\beq}{\begin{eqnarray}}
\newcommand{\eeq}{\end{eqnarray}}
\newcommand{\ist}{\;=\;}

\begin{document}

\title{
\begin{flushright}{\tiny UWThPh-2011-16}\end{flushright}
Revealing Bell's Nonlocality for Unstable Systems in High Energy Physics
}
%\author{Beatrix C. Hiesmayr$^1$, Antonio Di Domenico$^2$, Catalina Cureanu$^3$,  Andreas Gabriel$^1$, Marcus Huber$^1$, Jan-
%Ake Larsson$^4$, Pawel Moskal$^5$,...}
\author{Beatrix C. Hiesmayr}
\affiliation{University of
Vienna, Faculty of Physics, Boltzmanngasse 5, A-1090 Vienna, Austria}
\author{Antonio Di Domenico}
\affiliation{Sapienza Universit\`a di Roma, and INFN Sezione di Roma, Piazzale Aldo Moro 5, 00185 Rome, Italy}
\author{Catalina Curceanu}
\affiliation{Laboratori Nazionali di Frascati dell'INFN, 00044 Frascati, Italy}
\author{Andreas Gabriel}
\affiliation{University of
Vienna, Faculty of Physics, Boltzmanngasse 5, A-1090 Vienna, Austria}
\author{Marcus Huber}
\affiliation{University of
Vienna, Faculty of Physics, Boltzmanngasse 5, A-1090 Vienna, Austria}
\author{Jan-\r{A}ke Larsson}
\affiliation{Institionen f\"or Systemteknik, Linköpings Universitet, SE-58183 Linköping, Sweden}
\author{Pawel Moskal}
\affiliation{Institute of Physics, Jagiellonian University, Cracow, Poland}

%\date{Received: date / Revised version: date}

\begin{abstract}
Entanglement and its consequences - in particular the violation of Bell inequalities, which defies our concepts of realism and locality -
have been proven to play key roles in Nature by many experiments for various quantum systems. Entanglement can also be found in systems not consisting of ordinary matter and light,
i.e. in massive meson--antimeson systems. Bell inequalities have been discussed for these systems, but up to date no direct experimental test to conclusively exclude local realism was found. This mainly stems from the fact that one only has access to a restricted class of observables and that these systems are also decaying. In this Letter we put forward a Bell inequality for unstable systems which can be tested at accelerator facilities with current technology. Herewith, the long awaited proof that such systems at different energy scales can reveal the sophisticated ``\textit{dynamical}'' nonlocal feature of Nature in a direct experiment gets feasible. Moreover, the role of entanglement and $\mathcal{CP}$ violation, an asymmetry between matter and antimatter, is explored, a special feature offered only by these meson-antimeson systems.

\end{abstract}

%\keywords{Bell inequality, neutral kaons, foundations in Particle Physics}

\maketitle

\textbf{Introduction:}
The foundations of quantum mechanics have been extensively studied ever since the seminal work of Einstein, Podolsky and Rosen (EPR) in 1935, and the discovery of Bell inequalities~\cite{Bell} in 1964. Violations of the Bell inequalities, which reveal nonlocality of Nature, have been found in various distinct quantum systems~\cite{Aspect,Weihs,ion,Johsephsonqubits,Hasegawa}. Currently, more and more experiments in the realm of Particle Physics are exploring these issues~\cite{Hiesmayr3,Hiesmayr1,Hiesmayr13,GRW,Hiesmayr2,Bramon3,Bramon4,LIQiao,Genovese} which presently enter precision levels where, for various reasons, new physics is expected. In Refs.~\cite{Hiesmayr7,Hiesmayr8} and acknowledged in Refs.~\cite{HiesmayrKLOE,Aharonov} it was outlined that, in particular, the neutral K-meson system is suitable to show quantum marking and quantum erasure procedures in a way not available for ordinary matter and light. Therefore, this system is an exceptional laboratory for testing the very concepts of Nature. For mesonic systems one has two different measurement procedures, an \textit{active} one, exerting the free will of the experimenter, and a \textit{passive} one, with no control over the measurement basis nor on the time point.
For studies whether the strong correlations of the apparently paradoxical gedanken experiment by EPR can be explained by hidden parameters one has to demand that the two experimenters, commonly called Alice and Bob, independently and \textit{actively} choose among different alternatives. This rules out all meson systems except the neutral kaons whose sufficiently long lifetimes permit insertion of material at various places along their trajectories.

\begin{figure}
\centering
\includegraphics[width=6cm]{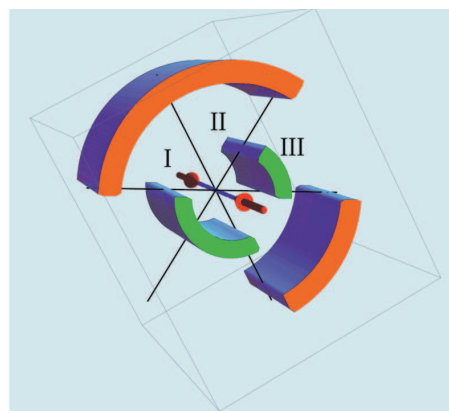}
\caption{{\bf Sketch of a possible setup for testing the Bell inequality.} The two beams collide in the origin and produce two neutral kaons propagating in opposite directions. Regions I,II,III cover measurements for different time choices ($\equiv$ distances). For any real experimental situation (e.g. pairs are not equally distributed in $4 \pi$) the geometry can be accordingly adapted.}\label{setup}
\end{figure}

There have been several proposals of Bell inequalities for the entangled kaonic system (e.g.Refs.~\cite{Hiesmayr3,Hiesmayr1,Hiesmayr13,GRW,Hiesmayr2,Bramon3,Bramon4,LIQiao}), but they lack a direct experimental verification, because both the observable space as well as the initial entangled state that can be produced at accelerator facilities is limited. These massive systems (about $1/2$ of a proton mass) are entangled in the quantum number strangeness, i.e. in being a particle and antiparticle, and present a unique laboratory to test for discrete symmetry violations as well as the foundations of quantum mechanics. Moreover, different to massive spin systems they transform trivially under the Lorentz group (see e.g. Refs.~\cite{rel1,rel2}), i.e. the entanglement is not observer dependent.

This letter starts by deriving a Bell inequality suited for decaying systems. Then we present a proposal how to test for violations of the Bell inequality with the KLOE-2 detector at the DA$\Phi$NE $e^+e^-$ collider of the Frascati Laboratory of INFN  (see e.g. Ref.~\cite{HiesmayrKLOE} and references therein) and discuss the experimental implementations followed by the analyses of limitations and loopholes.

\textbf{Bell Inequalities For Decaying Systems:}
In the EPR scenario a source produces two particles, which are separated and independently measured by Alice and Bob. Both parties can choose among two different measurements alternatives $i=n,n'$ for Alice and $j=m,m'$ for Bob. These settings yield either the outcomes $k,l=+1$ (later denoted as a yes event $Y$) or $k,l=-1$ (later denoted as a no event $N$).
Any classical or quantum correlation function can be defined in the usual way by
\begin{eqnarray}
E_{AB}(i,j)=\sum_{k,l} (k\cdot l)\; P_{AB}^{kl}(i,j)
\end{eqnarray}
where $P_{AB}^{kl}(i,j)$ is the joint probability for Alice obtaining the outcome $k$ and Bob obtaining the outcome $l$, when they chose measurements $i$ and $j$, respectively. For local realistic theories Bell's locality assumption imposes a factorization of the joint probabilities. Bell inequalities are tests for correlations that can be simulated using only local resources and shared randomness (a modern terminology for local hidden variables) and have, therefore, at hitherto nothing to do with quantum theory. Inserting the probabilities derived by quantum mechanics, however, in some cases leads  to a violation of the inequality, i.e. to a contradiction between predictions of local hidden variable theories and quantum theory. For bipartite entangled qubits a tight Bell inequality is the famous Clauser-Horne-Shimony-Holt (CHSH) Bell inequality~\cite{CHSH}, i.e.
\begin{widetext}
\begin{eqnarray}\label{chsh}
-2\leq S(n,m,n',m'):=E_{AB}(n,m)-E_{AB}(n,m')+E_{AB}(n',m)+E_{AB}(n',m')\leq 2\;.
\end{eqnarray}
In quantum mechanics the above inequality can also be rewritten in the so called witness form
\begin{eqnarray}\label{chshdecay}
\min_{\textrm{all}\;\rho_{sep}} S(n,m,n',m')[\rho_{sep}]\leq S(n,m,n',m')[\rho]\leq \max_{\textrm{all}\;\rho_{sep}}S(n,m,n',m')[\rho_{sep}]
\end{eqnarray}
where the extremum is taken over all separable states. The quantum mechanical correlations are derived by $E^{QM}_{AB}(n,m')(\rho)=Tr(O_n\otimes O_{m'}\rho)$ (where $O_i$ are appropriate operators) and hence the $S$-function can be rewritten by
\begin{eqnarray}
S(n,m,n',m')[\rho]=Tr(\biggl[O_n\otimes (O_m-O_{m'})+O_{n'}\otimes (O_m+O_{m'})\biggr]\;\rho)\;.
\end{eqnarray}
\end{widetext}
For stable systems the extremum over all separable states is always $2$, however, in case of unstable systems these bounds may become different from $2$ due to the decay property. From the above derivation it is obvious that the generalized Bell inequality holds for stable systems. In this case it is equivalent to the
famous CHSH-Bell inequality and all pure entangled states violate this inequality, whereas not all mixed entangled states directly lead to a violation.
The extremal Bell correlations for separable stable states always reach $2$, whereas for decaying systems we need to take into account the intrinsic decay property which as well affects the separable states. How and why this still constitutes a proper Bell inequality for decaying system will be elaborated in the discussion section.

In the following we present an experimental proposal to reveal the nonlocality given by the above generalized Bell inequality for entangled neutral K-mesons which are copiously produced at the DA$\Phi$NE accelerator facility. In these experiments the initially maximally entangled state is an antisymmetric singlet state. So far no Bell inequality was found for
\begin{itemize}
\item[(i)] \textit{active} measurements --a necessary requirement for a conclusive test \cite{Hiesmayr13}-- which leaves strangeness measurements as the only available basis choice without limitations and

\item[(ii)] the initial antisymmetric state, the only entangled state that is currently produced with sufficiently high luminosity.
\end{itemize}

 In particular it was shown \cite{Hiesmayr3,GRW,Hiesmayr2} that the CHSH-Bell inequality (\ref{chsh}) does not exceed the bounds $\pm2$ due to the fast decay compared to the oscillation in these kaonic systems. Thus, the reason for the non-violation is due to the given values of the two decay constants and the strangeness oscillation. In case the ratio of oscillation to decay would be twice as large~\cite{Hiesmayr13} then the quantum mechanical predictions would exceed the bounds $|2|$.  In Ref.~\cite{Hiesmayr1} it was shown that other initial states, in particular non--maximally entangled states, exceed the bounds $\pm2$, but up to date there is no experimental setup known that would produce such initial states.

\textbf{The Neutral Kaon System:}
In Ref.~\cite{Hiesmayr13} it was argued that any conclusive test against the existence of a local realistic description for entangled meson--antimeson pairs requires that Alice and Bob can choose among alternative measurements ``\textit{actively}''.
Particle detectors at accelerator facilities usually detect or reconstruct different decay products at various distances from the point of
generation, usually by a \textit{passive} measurement procedure, i.e. observing a certain decay channel at a certain position in the detector without having
control over the decay channel nor the time (determined by the distance from point of generation). More rarely an \textit{active} measurement procedure is performed, e.g. by placing a
piece of matter in the beam and forcing the incoming neutral meson to interact with the material (see e.g. the 1998 CPLEAR experiment~\cite{CPLEAR}). For practical reasons (too big decay constants) this procedure is only possible for neutral K-mesons and not for the neutral $B$ and $D$ mesons. In such a way the strangeness content of a neutral kaon, i.e. being
a kaon $K^0$ or an antikaon $\bar K^0$, at a certain time (determined by the distance of the piece of matter from the source) can be measured \textit{actively}, i.e. the experimenter decides what physical property (in this case strangeness) is measured \textbf{and} when she or he wants to measure. Neutral kaons have rather long lifetimes enabling them to travel several centimeters or even meters (depending on their velocities), therefore kaonic qubits present a quantum system that is entangled over macroscopic distances. Certainly, for any conclusive test of a Bell inequality it is necessary that Alice and Bob can freely choose among different options. This requirement, which has been stated by the authors of Ref.~\cite{Hiesmayr13}, is not a loophole, and therefore rules out all the other meson systems.

In principle, one can generally ask the following dichotomic questions to the unstable quantum systems
\begin{itemize}
\item[(i)] Are you in the quasispin state $|k_n\rangle$, i.e. a certain superposition of the two strangeness eigenstates $|K^0\rangle$ and $|\bar K^0\rangle$, at a certain time $t_n$ or not? (Answers: $Yes (Y)/No (N)$)
\item[(ii)] Are you in the quasispin state $|k_n\rangle$ or its orthogonal state $|k_n^\perp\rangle$ ($\langle k_n^\perp|k_n\rangle=0$) at a certain time $t_n$?
\end{itemize}
The second question $(ii)$ does not include all available information of the unstable quantum systems under discussion as it ignores cases where the neutral kaons decayed before the question was asked at time $t_n$. Again, it is crucial for any conclusive test of a Bell inequality not to ignore available information, thus we have to stick to the first question.

In a recent publication \cite{Heisikaon} an effective formalism was developed for expressing the quantum mechanical expectation value by effective time dependent operators in the reduced Hilbert--Schmidt space of the surviving component. This has the advantage that the Bell inequality can be formulated as a witness operator and, for a given initial state, the value of the Bell inequality can be simply derived. Mathematically, finding out whether a Bell inequality is violated is a highly constrained optimization problem even for qudit systems. In detail, any quantum expectation value for any choice of measurements is given by an effective $2\times2$ operator in the Heisenberg picture for a given initial state $\rho$ (not necessarily pure):
\begin{widetext}
\begin{eqnarray}
&&E^{QM}(k_n,t_n; k_m,t_m)\ist Tr (O_n\otimes O_m\;\rho)\nonumber\\
&&\;=\;P(\textrm{Y}: k_n,t_n;\textrm{Y}: k_m,t_m)+P(\textrm{N}: k_n,t_n;\textrm{N}: k_m,t_m)%\nonumber\\
- P(\textrm{Y}: k_n,t_n;\textrm{N}: k_m,t_m)
-P(\textrm{N}: k_n,t_n;\textrm{Y}: k_m,t_m)\;,
%&\stackrel{P(\textrm{Yes}: k_n,t_n)+P(\textrm{No}: k_n,t_n)=1}{=}&2\; P(\textrm{Yes}: k_n,t_n)-1\nonumber\\
%&=&
%\ist Tr (O_n\otimes O_m\;\rho)\nonumber
\end{eqnarray}
where $P(\textrm{Y/N},\textrm{Y/N})$ denote the joint probabilities and $O_n\;:=\;\lambda_n\;|\chi_n\rangle\langle\chi_n|-|\chi_n^\perp\rangle\langle\chi_n^\perp|$ is an effective $2\times 2$ operator with $\langle \chi_n|\chi_n^\perp\rangle\ist 0$, $N(t_n)\ist e^{-\Gamma_S t_n}\; |\langle K_S|k_n\rangle|^2+e^{-\Gamma_L t_n}\; |\langle K_L|k_n\rangle|^2$ and
\begin{eqnarray}\label{op}
|\chi_n\rangle&=&
\frac{1}{\sqrt{N(t_n)}}
\biggl\lbrace \langle K_S|k_n\rangle\cdot e^{(i m_S-\frac{\Gamma_S}{2}) t_n}\;|K_1\rangle+\langle K_L|k_n\rangle\cdot e^{(i m_L-\frac{\Gamma_L}{2}) t_n}\;|K_2\rangle\biggr\rbrace\nonumber\\
\lambda_n&=&-1+ (e^{-\Gamma_S t_n}-e^{-\Gamma_L t_n}) (1-\delta^2) \cos\theta_n+ (e^{-\Gamma_S t_n}+e^{-\Gamma_L t_n})(1+\delta^2+2\delta \cos\phi_n \sin\theta_n)\;.\nonumber
\end{eqnarray}
\end{widetext}
Here we parameterized the quasispin $k_n$ as a superposition of the $\mathcal{CP}$ eigenstates $|K_{1/2}\rangle$, i.e. $|k_n\rangle\ist \cos\frac{\theta_n}{2}|K_1\rangle+\sin\frac{\theta_n}{2}\cdot e^{i \phi_n}\;|K_2\rangle$. The eigenstates $|K_{S/L}\rangle$, i.e. the short lived state $|K_S\rangle$ and the long lived state $|K_L\rangle$, are the mass eigenstates which are the solutions of the effective Schr\"odinger equation. $\Gamma_{S,L}$ are their decay constants and $\Delta m=m_L-m_S$ is the mass difference. The parameter $\delta$ is defined as $\delta=\frac{2 \Re \epsilon}{1+|\epsilon|^2}$, where $\epsilon$ is
the small $\mathcal{CP}$ violating parameter $O(10^{-3})$, i.e. quantifying the asymmetry between a world of matter and antimatter. Note that due to the decaying property of the system the first eigenvalue $\lambda_n$ changes in time depending on the measurement choice and approaches the value $-1$ for $t_n\longrightarrow\infty$, independently of the choice of the observer.

For spin--$\frac{1}{2}$ systems, the most general spin observable is given by $O_n\equiv\vec{n}\vec{\sigma}$ with the Pauli operators $\sigma_i$. Here any normalized quantization direction ($|\vec{n}|=1$) can be parameterized by the azimuth angle $\theta_n$ and the polar angle $\phi_n$. In case of unstable systems the effective observable is given by the set of operators $O_n\equiv\frac{\lambda_n-1}{2}\mathbbm{1}+\frac{\lambda_n+1}{2}\vec{n}(\theta_n,\phi_n,t_n)\vec{\sigma}$ for which the ``\textit{quantization direction}'' is no longer normalized and its loss results in an additional contribution in form of ``white noise'', i.e. the expectation value gets a contribution independent of the initial state for $t_n>0$.

\textbf{Experimentally Testable Bell Inequality and Experimental Feasibility:}
Consider the usual EPR scenario of a source emitting entangled pairs of particles, in case of the DA$\Phi$NE collider the decay of a $\Phi$-meson into two neutral kaons in the antisymmetric maximally entangled Bell state at time $t=0$, $|\psi^-\rangle=\frac{1}{\sqrt{2}}\lbrace |K^0 \bar K^0\rangle-|\bar K^0 K^0\rangle\rbrace$. Alice and Bob agree to measure the strangeness content actively (we choose e.g. `\textit{Are you a $\bar K^0$ or not}'), i.e. by inserting a piece of material in the beam of neutral kaons. Both choose fully randomly and independently at what time they measure the neutral kaons (distances from the source). A possible setup is sketched in Fig.~\ref{setup}.

\begin{figure}
\centering
(a)\includegraphics[keepaspectratio=true, width=8cm]{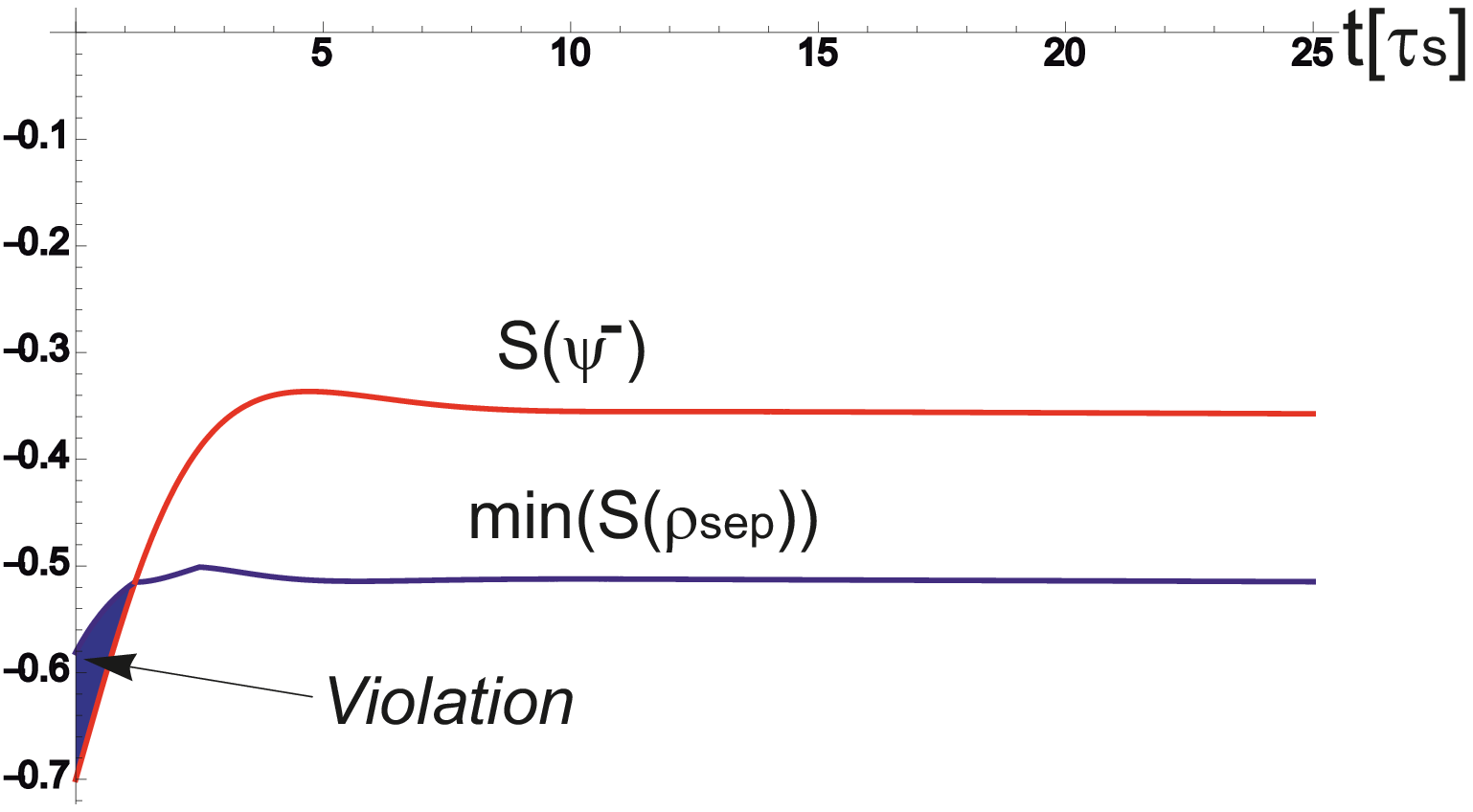}
(b)\includegraphics[keepaspectratio=true, width=8cm]{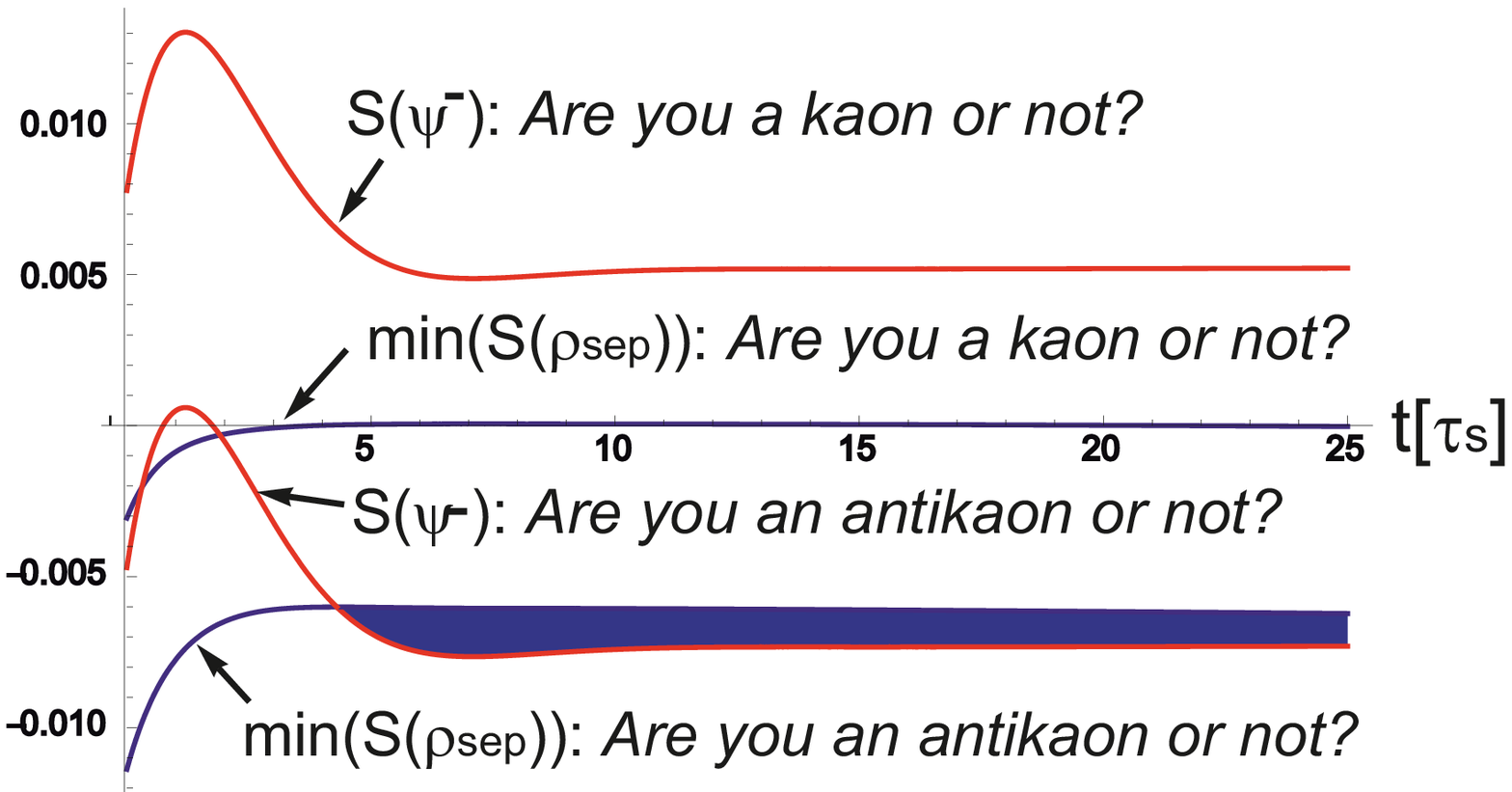}
\caption{{\bf The generalized CHSH-Bell inequality for decaying systems.} These graphs show the functions $S(|\psi^-\rangle)$ and $\min S(\rho_{sep})$ for the time choices (a)  $t_m=t_{n'}=1.34\tau_S, t_{m'}=2.80\tau_S$ varied over $t_n=t$ and (b) $t_n=4.48\tau_S,t_m=t_{n'}=4.81\tau_S$ varied over $t_{m'}=t$ in the units of $\tau_S$. The violation in (b) vanishes for $t_n\geq 80 \tau_S$.}\label{figviolation}
\end{figure}

One example for a choice of the four involved times is e.g. $t_n=0, t_m=t_{n'}=1.34\tau_S, t_{m'}=2.80\tau_S$ ($\tau_S$\dots lifetime of the short lived state)  which leads to $S(|\psi^-\rangle)=-0.69$ and $\min S(\rho_{sep})=-0.58$, thus a violation of $0\leq\Delta:=S(|\psi^-\rangle)-\min S(\rho_{sep})=-0.11$. Certainly, a measurement at $t_n=0$ is not possible, but it can be increased up to $t_n=1.34 \tau_S$ as visualized in Fig.~\ref{figviolation}~(a).

Choices of times with higher values also yield violations, as visualized in Fig.~\ref{figviolation}~(b), however, these are due to the small $CP$ violation parameter and the big difference of the decay rates. Here the question raised to the system, i.e. ``\textit{Are you an antikaon or not?}'' or ``\textit{Are you a kaon or not?}'' matters both for antisymmetric state as well as for the lower bound derived for all possible separable states. Consequently, this means that the interference caused by $\mathcal{CP}$ violation can as well reveal the nonlocality of this system.

This peculiar relationship between a symmetry violation in Particle Physics and manifestation of entanglement can also be derived when one chooses a Bell inequality varied in the quasispins. In particular, there exists a set of Bell inequalities~\cite{Hiesmayr3} for the antisymmetric Bell state that require the $CP$ violation parameter $\delta=0$, i.e. local hidden variable theories are in contradiction to the measured asymmetry of matter and antimatter. This puzzling relation between symmetry violations in Particle Physics and manifestations of entanglement can not be put to a direct experimental verification due to technological limitations, different to the Bell inequality proposed in this letter.

\textbf{Discussion:}
In any experiment testing a Bell inequality one faces loopholes, i.e. has to make supplementary assumptions. There are two prominent loopholes. The first one is called the `detection loophole', stating that if not all pairs are measured or if some are misidentified due to imperfections of the detectors, Nature could still be local since some information is missing.  This loophole affects especially photon experiments as even the best available detectors only detect a fraction of the pairs~\cite{Aspect,Weihs}, thus these experiments rely on the `fair sampling' hypothesis. The second loophole, the `locality loophole', states that measurements of Alice and Bob have to be space-like separated, thus avoiding any possible exchange of subluminal signals about the measurement choices of Alice and Bob. The `detection loophole' was closed with experiments on beryllium ions~\cite{ion} and Johsephson phase qubits~\cite{Johsephsonqubits} while the `locality loophole' was closed for photons~\cite{Weihs}, but up to date no experiments exist closing both loopholes simultaneously, but there are several proposals, e.g. Ref.~\cite{Rosenfeld}.

It was claimed in the beginning that these massive entangled systems which are copiously produced and separated into opposite directions with relativistic velocities could offer a possibility of simultaneously closing both mentioned loopholes, however, as e.g. intensively discussed in Ref.~\cite{BramonEscribanoGabarino} the real situation is far more involved. Moving materials very close to the beam inside a particle detector would cause serious problems as it would have many experimental side effects hard to control and would influence other measurements. Also having ``static'' material very close to the beam  is a challenge since it could change the beam performance, however, it is conceivable and feasible to design such an experimental facility, e.g. by exploiting the thin cylindrical pipe where the beams circulate (in the KLOE-2 detector the pipe radius is $3.7$cm corresponding to about $6 \tau_S$).  Note that the efficiency of the required strangeness measurements is less than naively expected from the strong nature of these interactions (see e.g. Ref.~\cite{DiDomenico}). The difficulty does not stem from detecting the reaction products but rather from the low  probability in initiating the strong reaction in a thin slab of material.

This has to be taken into account when counting the $Yes$ and $No$-events, i.e. in correctly evaluating the detection efficiency. This difficult task can be addressed with the help of the Monte Carlo simulation of the KLOE-2 detector, which carefully takes into account the well studied performance of the collider and detector. Additional checks on independent samples of experimental data (e.g. comparing the detected numbers of $K^0$ and $\bar K^0$ with pure $K_{S/L}$ beams) give another good experimental test in order to control the errors.

Further advantages of these decaying systems is that one knows essentially with $100\%$ probability that in case a neutral kaon is reconstructed it can only come from an entangled pair. In addition, on average only one entangled pair is generated per event. All that provides a very clean environment and gives high precision in measuring the joint and single probabilities, respectively. Therefore, the expectation values can be also evaluated by measuring only joint and single probabilities, differently from experiments with photons, which usually rely on coincidence counts. Consequently, one can as well test the Clauser-Horne (CH) version~\cite{ClauserHorn} of the CHSH-Bell inequality which requires single and joint probabilities. The fundamental difference between the CHSH-Bell inequality and the CH-Bell inequality stems from the fact that in the first case correlations based on only joint probability measurements are tested whereas the other one involves only probabilities (single and joint ones).

For stable systems the extremal Bell correlations are always achieved for pure states due to convexity of the expectation value of the Bell operator. Due to the unavoidable decay the extremal Bell correlations can be significantly lower as in any measurement basis the probability to obtain a Yes-event becomes distorted. In contrast to the detection loophole in our setup we have full control over all joint and single probabilities and the full account of all decay events. This is significantly different to previous proposals, e.g. Ref.~\cite{GisinGo}, where each Bell correlation was normalized to surviving pairs, i.e. the question to the system corresponds to ``\textit{Are you an antikaon or a kaon at time $t$?}'' (question type (ii)). This means that all pairs that did not survive until the measurement times are discarded, clearly not testing the whole ensemble. Consequently, our generalized Bell inequality is a conclusive test of Bell's nonlocality under the assumption that the time evolution (exponential decay) of \textit{single} kaons is correctly described by quantum mechanics. Obviously, a local realistic theory has not to obey any quantum laws, but it is natural to demand that any local realistic theory also predicts all measurable single probabilities correctly. This is what is taken into account via our extremal bounds.

In Ref.~\cite{Selleri} the authors proposed quite general local realistic models for the antisymmetric Bell state and measurements of antikaons at different times (however, not incorporating $\mathcal{CP}$ violation). The models assume that the time evolution of the single kaon predictions are correct, i.e. those of quantum mechanics. We adapted their model to compare it with our generalized Bell inequality. We find that the lower and upper bounds of their models are less or equally stringent than our bounds, i.e. our generalized Bell inequality provides a more stringent test of nonlocality.

Finally, let us mention that the obtained violation strongly depends on the difference of the decay constants of the short and long lived states which is a special feature of the neutral kaon system (for all other meson-antimeson systems it is essentially zero).

\textbf{Conclusion:}
In conclusion, the proposed generalized Bell inequality for decaying systems for restricted observable space and dichotomic questions submits to test our
conception of locality and reality in a \textit{dynamical} way including $\mathcal{CP}$ violation, whose origin is still a big puzzle in Physics.
It presents the first conclusive test, i.e. does not fail due to unavoidable requirements, but involves loopholes. Moreover, a direct experimental test for the antisymmetric maximally entangled Bell state produced at accelerators become possible. In particular, a test with the KLOE-2 detector at the DA$\Phi$NE collider is feasible. Even if our proposal at a first step is realized with a static measurement setup and in the presence of other loopholes, it is a step towards proving the peculiar consequences of entanglement for massive systems at different realms of energy. Herewith, it also contributes to the open question which role entanglement and $\mathcal{CP}$ violation plays in our universe.

\textbf{Acknowledgements}
The authors B.C.H., A.G. and M.H. want to thank the Austrian Science Fund (FWF) project P21947N16. B.C.H and C.C. thank also the COST action MP 1006.

%\textbf{Author Contributions}
%B.C.H. derived the proposal for the generalized Bell inequality for unstable systems, A. D.D. included the details for the experimental realization, the other authors (added in alphabetical order) equally checked the result and contributed to the discussions.

\end{document}